\documentclass{PoS}
\usepackage{amsmath}
\title{Searches for strong dynamics signals in gauge boson pair production}

\ShortTitle{Searches for strong dynamics signals in gauge boson pair production}

\author{\speaker{Oscar Cat\`a}\thanks{This work was performed in the context of the ERC Advanced Grant project 'FLAVOUR' (267104) and was supported in part by the DFG cluster of excellence 'Origin and Structure of the Universe'.}\\
        Ludwig-Maximilians-Universit\"at M\"unchen, Fakult\"at f\"ur Physik,
Arnold Sommerfeld Center for Theoretical Physics, 
D--80333 M\"unchen, Germany
\\
        E-mail: \email{oscar.cata@physik.uni-muenchen.de}}

\abstract{I briefly review the construction of the effective field theory of a nonlinearly realized electroweak symmetry breaking and apply it to the study of gauge boson pair production. I will consider $WW$, $ZZ$ and $\gamma Z$ production at linear colliders as reference illustrative examples. I will show that in all cases a consistent effective field theory treatment allows to encode the dominant new physics effects entirely as gauge-fermion vertex corrections. From naive dimensional analysis one expects these corrections to be at the per mille level. However, they grow with energy, such that at TeV colliders (LHC and futures linear collider facilities) these new physics effects can easily amount to 20$\%$ corrections to the production cross sections.}

\FullConference{The 7th International Workshop on Chiral Dynamics,\\
		August 6 -10, 2012\\
		Jefferson Lab, Newport News, Virginia, USA}

\begin{document}
\section{Introduction}
Gauge boson pair production is one of the simplest processes to test the triple-gauge structure predicted by the standard model. Based on gauge and Lorentz invariance, general corrections to the vertices can be cast in terms of 7 kinematic invariants for the charge diboson production~\cite{Hagiwara:1986vm}: 
\begin{align}
\Gamma_{\mu\nu\lambda}^{WWV}(p_1,p_2;q)&=f_1Q_{\lambda}g_{\mu\nu}+f_2(p_{1\nu}g_{\mu\lambda}-p_{2\mu}g_{\nu\lambda})+if_3(p_{1\nu}g_{\mu\lambda}+p_{2\mu}g_{\nu\lambda})+f_4p_{1\nu}p_{2\mu}Q_{\lambda}\nonumber\\
&+if_5\epsilon_{\mu\nu\lambda\rho}Q^{\rho}+f_6\epsilon_{\mu\nu\lambda\rho}q^{\rho}+f_7Q_{\lambda}\epsilon_{\mu\nu\alpha\rho}p_{1}^{\alpha}p_{2}^{\rho}
\end{align}
where $Q=p_1-p_2$. For the neutral case, gauge invariance and Bose symmetry further reduce the number of parameters to   
\begin{align}\label{ZZ}
\Gamma_{\mu\nu\lambda}^{ZZV}(p_1,p_2;q)&=i\mu_1^V(p_{1\nu}g_{\mu\lambda}+p_{2\mu}g_{\nu\lambda})+i\mu_2^V\epsilon_{\mu\nu\lambda\rho}Q^{\rho}
\end{align}
and
\begin{align}
\Gamma_{\mu\nu\lambda}^{Z\gamma V}(p_1,p_2;q)&=i\beta_1^V\left(p_{2\lambda}g_{\mu\nu}-p_{2\mu}g_{\nu\lambda}\right)+i\beta_2^V\epsilon_{\mu\nu\lambda\rho}p_{2}^{\rho}\nonumber\\
&+i\frac{\beta_3^V}{\Lambda^2}p_{2\mu}(p_1\cdot p_2g_{\nu\lambda}-p_{1\nu}p_{2\lambda})+i\frac{\beta_4^V}{\Lambda^2}Q_{\lambda}\epsilon_{\mu\nu\alpha\beta}p_1^{\alpha}p_2^{\beta}
\end{align}
Most of the existing analyses on gauge diboson production rely on the assumption that triple-gauge corrections are the dominant source of new physics effects. However, from the global LEP electroweak fit there is no evidence that other new physics contributions, namely gauge-boson corrections, be suppressed~\cite{Han:2004az}. Actually, at least for the charged diboson production, gauge-fermion corrections are a priori expected to compete with triple-gauge ones. In order to improve and extend the scope of the analysis to cover all possible new physics sources, an effective field theory (EFT) seems the only consistent tool. An EFT selects the relevant physics at a given scale while retaining field theory symmetries. This implies that, as opposed to traditional form-factor analyses, in an EFT approach the triple-gauge parameters above will automatically comply with the standard model symmetries. 

Here I will work under the assumption that the dynamics underlying electroweak symmetry breaking (EWSB) are of strongly-coupled nature at the TeV scale. The most general EWSB pattern with the minimal particle content is $SU(2)_L\times SU(2)_R\to SU(2)_V$, which generates 3 Goldstone bosons (the longitudinal modes of the W and Z) collected in the nonlinear matrix $U\to g_LUg_R^{\dagger}$. In this scenario, a light scalar field can be added as a singlet~\cite{Contino:2010mh}. 

Over the years, there have been several EFT-inspired analysis of gauge boson pair production. However, a full-fledged EFT analysis was lacking, mostly because the systematics of the associated power-couting were unclear. These issues were clarified in~\cite{Buchalla:2012qq} and the full characterization of NLO operators was given. Subsequent applications of the formalism were given for charged~\cite{Buchalla:2013wpa} and neutral~\cite{Cata:2013sva} gauge boson pair production. In the following I will briefly review the content of those three papers. I will show that the interplay of EFT techniques and the high-energy window accessible at LHC and linear colliders provides a rather simple picture for the new physics searches in $WW$, $ZZ$ and $\gamma Z$ production. 

\section{EWSB from strongly-coupled dynamics}
When EWSB is originated from strongly-coupled interactions, the standard model Lagrangian can be seen as a nonlinearly-realized effective theory, valid up to a dynamically-generated new physics scale $\Lambda\sim 4\pi v\sim 3$ TeV, whose leading-order term is
\begin{align}
{\cal L}_{LO}&=-\frac{1}{2}\langle W_{\mu\nu}W^{\mu\nu}\rangle 
-\frac{1}{4} B_{\mu\nu}B^{\mu\nu}+i\sum_j\bar f_j \!\not\!\! Df_j+\frac{v^2}{4}\ \langle D_\mu U D^\mu U^{\dagger}\rangle+{\cal{L}}_{Yukawa}(U,{\bar{f}},f)
\end{align}
One can show that the degree of divergence $\Delta$ of each diagram generated from ${\cal{L}}_{LO}$ is given by~\cite{Buchalla:2012qq}
\begin{align}
\Delta=v^2(yv)^{\nu_f}(gv)^{\nu_g}\frac{p^d}{\Lambda^{2L}}\left(\frac{\Psi}{v}\right)^F\left(\frac{X_{\mu\nu}}{v}\right)^G \left(\frac{\varphi}{v}\right)^B
\end{align} 
where $d=2L+2-F/2-G-\nu_f-\nu_g$ (see~\cite{Buchalla:2012qq} for details). Since $d$ is bounded from above, this leads to a consistent power-counting, {\it{i.e.}} the number of counterterms at any order is finite. With $U,\psi,X$ as shorthand notations for Goldstone, fermion and gauge fields, one can show that at NLO there are 6 classes of operators, denoted as $UD^4$, $XUD^2$, $X^2U$, $\psi^2UD$, $\psi^2UD^2$ and $\psi^4U$, while at NNLO one finds $UD^6$, $UXD^4$, $X^2UD^2$, $\psi^2 UD^3$, $X^3U$, $\psi^2 UX$ and $\psi^2 UXD$.

\section{New physics in gauge diboson production}

\begin{figure}[t]
\begin{center}
\includegraphics[width=2.5cm]{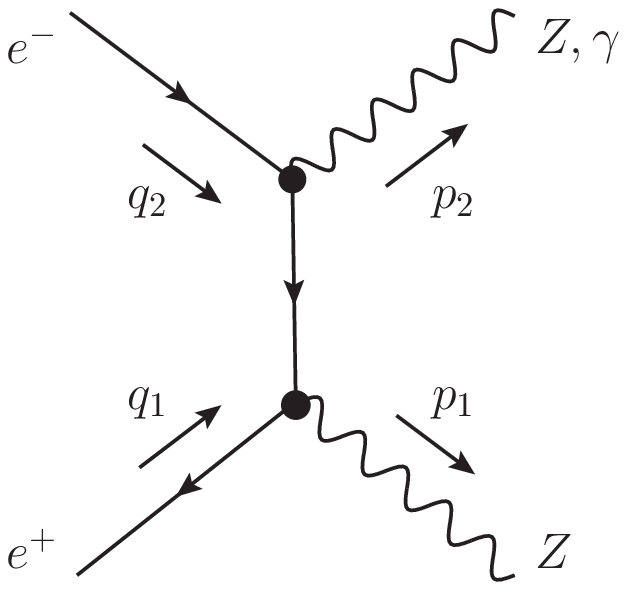}
\hspace{0.3cm}
\includegraphics[width=2.5cm]{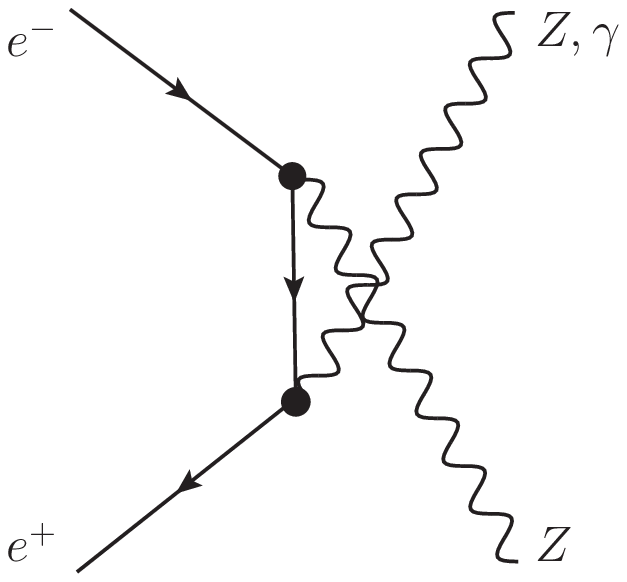}
\hspace{0.3cm}
\includegraphics[width=2.9cm]{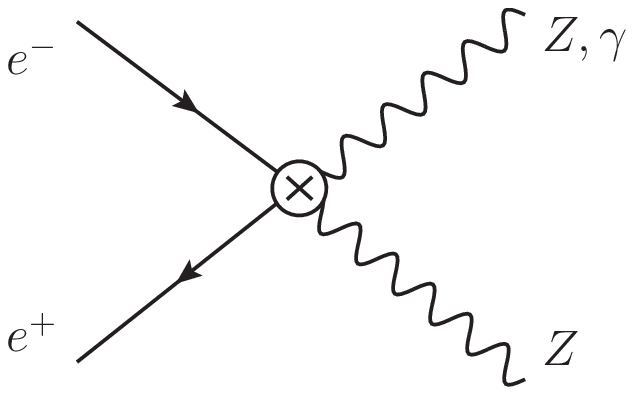}
\hspace{0.3cm}
\includegraphics[width=4.3cm]{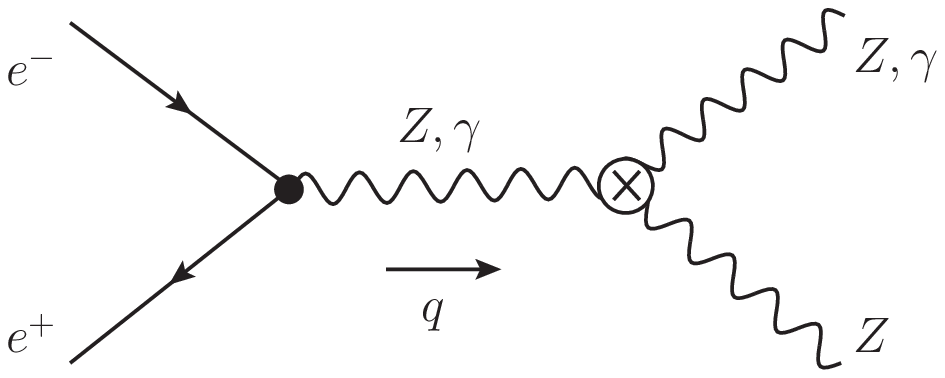}
\end{center}
\caption{\small{\it{Different topologies contributing to neutral gauge boson production. Dotted vertices start at leading order, while crossed ones appear at NNLO.}}}\label{fig:1}
\end{figure}

As an application, one can consider the new physics effects in ${\bar{f}}f\to WW,\gamma Z,ZZ$. In the unitary gauge, the leading new physics in $WW$ production can be described diagrammatically as corrections to the $s$ and $t$-channel standard model (SM) tree level contributions. For the neutral case, instead, corrections affect the $t,u$-channel SM diagrams, but also generate $s$-channel and contact term contributions (see Fig~\ref{fig:1}). These {\emph{direct}} contributions to gauge-fermion and triple-gauge vertices however do not exhaust new physics effects, which also shift the gauge boson kinetic terms and the fundamental standard model parameters $m_Z,\alpha_{em},G_F$. At NLO, the relevant subset of operators for the direct and indirect contributions comes from the classes $X^2U$, $\psi^2UD$ and $\psi^4U$. It reads
\begin{align}
{\cal{L}}_{NLO}&=\sum_j^6 \lambda_j {\cal{O}}_{Xj}+\sum_j \eta_j {\cal{O}}_{Vj}+\beta{\cal{O}}_{\beta}+\eta_{4f}{\cal{O}}_{4f}
\end{align}
where ${\cal{O}}_{\beta}=v^2\langle \tau_LL_{\mu}\rangle^2$ and
\begin{align}\label{naive1}
{\cal{O}}_{X1}&=g^{\prime}gB_{\mu\nu}\langle W^{\mu\nu}\tau_L\rangle\qquad &&{\cal{O}}_{X4}=g^{\prime}g\epsilon^{\mu\nu\lambda\rho}\langle \tau_L W_{\mu\nu}\rangle 
B_{\lambda\rho}\nonumber\\
{\cal{O}}_{X2}&=g^2 \langle W^{\mu\nu}\tau_L\rangle^2 \qquad&&{\cal{O}}_{X5}=g^2\epsilon^{\mu\nu\lambda\rho}\langle\tau_LW_{\mu\nu}\rangle
\langle\tau_LW_{\lambda\rho}\rangle\nonumber\\
{\cal{O}}_{X3}&=g\epsilon^{\mu\nu\lambda\rho}\langle W_{\mu\nu}L_{\lambda}\rangle\langle\tau_L L_{\rho}\rangle\qquad &&{\cal{O}}_{X6}=g\langle W_{\mu\nu}L^{\mu}\rangle\langle \tau_LL^{\nu}\rangle
\end{align}
are oblique and triple-gauge corrections,
\begin{align}
{\cal O}_{V1}&=-\bar q\gamma^\mu q\ \langle L_{\mu}\tau_L\rangle; \qquad &&{\cal O}_{V7}=-\bar l\gamma^\mu l\ \langle L_{\mu}\tau_L\rangle\nonumber\\
{\cal O}_{V2}&=-\bar q\gamma^\mu \tau_L q\ 
\langle L_{\mu}\tau_L\rangle;\qquad && {\cal O}_{V8}=-\bar l\gamma^\mu \tau_L l\ 
\langle L_{\mu}\tau_L\rangle\nonumber\\
{\cal O}_{V4}&=-\bar u\gamma^\mu u\ \langle L_{\mu}\tau_L\rangle;\qquad && {\cal O}_{V9}=-\bar l\gamma^\mu \tau_{12} l\ 
\langle  L_{\mu}\tau_{21}\rangle\nonumber\\
{\cal O}_{V5}&=-\bar d\gamma^\mu d\ \langle L_{\mu}\tau_L\rangle;\qquad && {\cal O}_{V10}=-\bar e\gamma^\mu e\ \langle L_{\mu}\tau_L\rangle
\end{align}
are gauge-fermion new physics contributions and
\begin{align}
{\cal{O}}_{4f}&=\frac{1}{2}({\cal{O}}_{LL5}-4{\cal{O}}_{LL15})=({\bar{e}}_L\gamma_{\rho}\mu_L)({\bar{\nu}}_{\mu}\gamma^{\rho}\nu_e)
\end{align}
Above, $L_{\mu}=iUD_{\mu}U^{\dagger}$, $\tau_L=\displaystyle UT_3U^{\dagger}$, $\tau_{12}=T_1+iT_2$ and $\tau_{21}=T_1-iT_2$ have been used as shorthand notation. ${\cal{O}}_{X1,2}$, ${\cal{O}}_{\beta}$, ${\cal{O}}_{V9}$ and ${\cal{O}}_{4j}$ are indirect contributions: the first two shift the kinetic terms, the third renormalizes $m_Z$ and the remaining two renormalize $G_F$. ${\cal{O}}_{V1-5,9}$ are the 5 direct contributions to gauge-fermion vertices in $pp$ collisions (correcting ${\bar{u}}d W$, ${\bar{u}}uZ$ and ${\bar{d}}dZ$ vertices), while ${\cal{O}}_{V9,10}$ and the combination $\frac{1}{2}{\cal{O}}_{V7}-{\cal{O}}_{V8}$ correct the ${\bar{e}}eZ$ and ${\bar{\nu}}eW$ vertices.

The best strategy is to eliminate the indirect contributions by field and parameter redefinitions~\cite{Holdom:1990xq}. By canonically normalizing gauge kinetic and SM parameters, the net effect is a shift in the gauge-fermion vertices  
\begin{align}
{\cal{L}}_f&=e{\bar{f}}\gamma_{\mu}A^{\mu}f+e\sum_{j=L,R}\bigg[\zeta_j^{(0)}+\delta\zeta_j\bigg]{\bar{f}}_j\gamma_{\mu}Z^{\mu}f_j-\frac{g}{\sqrt{2}}\bigg[1+\delta\phi_L\bigg]{\bar{\nu}}_L\gamma^{\mu}W_{\mu}^+f_L+{\mathrm{h.c.}}
\end{align} 
where $\zeta_L^{(0)}=t_{2W}^{-1}$ and $\zeta_R^{(0)}=-t_{W}$ are the standard model tree level values and  
\begin{align}\label{gfEFT}
\delta \zeta_L=\frac{2e^2\lambda_1-c_{2W}\eta_L+{\hat{\beta}}}{s_{2W}c_{2W}};\qquad
\delta \zeta_R=\frac{2e^2\lambda_1-c_{2W}\eta_R+2s_W^2{\hat{\beta}}}{s_{2W}c_{2W}}
\end{align}
$\eta_{L,R}$ collect the full left and right-handed contributions to $Z$ gauge-fermion corrections. A similar expression holds for $\delta\phi_L$~\cite{Buchalla:2013wpa}.

For $WW$ production, ${\cal{O}}_{X1-6}$ appear as direct contributions to the triple-gauge vertices. In contrast, for $ZZ$ and $\gamma Z$ production, there is no triple-vertex contribution at NLO and instead one has to go to NNLO. There are 8 operators in $\psi^2UXD$ that contribute to neutral triple-gauge vertices ($J_{\mu}^{(L,R)}=({\bar{f}}\gamma_{\mu}f)_{L,R}$)~\cite{Cata:2013sva}: 
\begin{align}\label{basis}
{\cal{L}}_{nTGV}=\sum_{j=L,R}\!\!\bigg\{\frac{c_{jW}}{\Lambda^2}J_{\mu}^{(j)}\langle W^{\mu\nu}L_{\nu} \rangle+\frac{c_{jB}}{\Lambda^2}J_{\mu}^{(j)} B^{\mu\nu} \langle \tau_L L_{\nu} \rangle+\frac{{\tilde{c}}_{jW}}{\Lambda^2}J_{\mu}^{(j)}\langle {\tilde{W}}^{\mu\nu}L_{\nu} \rangle+\frac{{\tilde{c}}_{jB}}{\Lambda^2}J_{\mu}^{(j)} {\tilde{B}}^{\mu\nu} \langle \tau_L L_{\nu} \rangle\bigg\}
\end{align}

\section{Signatures at LHC and linear colliders}
The existence of a large energy gap between the electroweak and new physics scales is precisely the motivation to adopt an EFT viewpoint. In this work we are aiming at the energy window $\sqrt{s}\sim (0.6-1)$ TeV, which can be easily achieved at the LHC and linear colliders. In this regime $v^2\ll s\ll \Lambda^2$ and consequently (i) an expansion of the cross sections in powers of $v^2/s$ is allowed; and (ii) good convergence of the EFT expansion (in $s/\Lambda^2$) is expected.

\begin{figure}[t]
\begin{center}
\includegraphics[width=4.5cm]{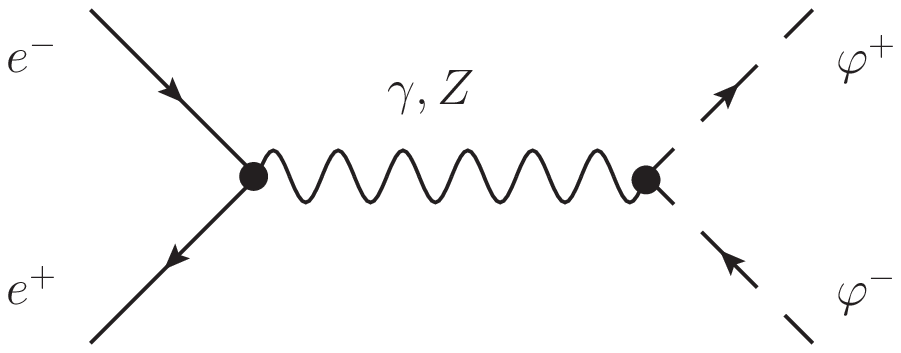}\hskip 0.5cm
\includegraphics[width=2.8cm]{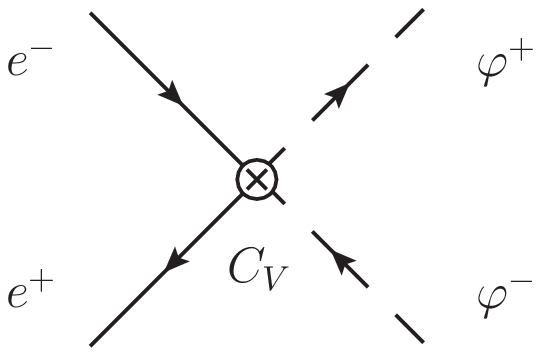}\hskip 0.5cm
\includegraphics[width=4.5cm]{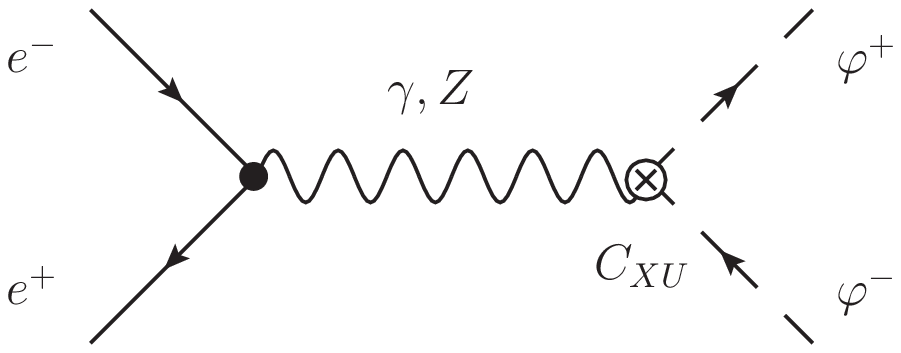}
\end{center}
\caption{\small{\it{Different contributions to $e^+e^-\to \varphi^+\varphi^-$. From left to right: (i) Standard Model piece; new physics contribution in terms of (ii) gauge-fermion operators and (iii) triple-gauge operators.}}}\label{fig:2}
\end{figure}

In this fiducial regime, the interference between the Standard Model and the new physics contribution in $WW$ production can be easily computed and results in~\cite{Buchalla:2013wpa}
\begin{align}\label{results}
\frac{d\sigma^R_{WW}}{d\cos\theta}=\frac{\pi \alpha^2 \sin^2\theta}{8 s_W^2c_W^2}\frac{1}{m_W^2}\eta_R;\qquad 
\frac{d\sigma^L_{WW}}{d\cos\theta}=\frac{\pi \alpha^2 \sin^2\theta}{16 c_W^2s_W^4}\frac{1}{m_W^2}\eta_{L}
\end{align}
which can be entirely cast in terms of gauge-fermion new physics operators. Working out the different final-state polarizations, one actually learns that the previous results come entirely from the $W_LW_L$ production. The result can be best understood by looking at Goldstone boson production in Landau gauge (see Fig.~\ref{fig:2}). New physics contributions come from the contact terms (central diagram), which are generated by the gauge-fermion operators.

A key observation to obtain Eq.~(\ref{results}) is that commonly used triple-gauge operators: 
\begin{align}
{\cal{O}}_{X7}=ig^{\prime}B_{\mu\nu}\langle\tau_L[L^{\mu},L^{\nu}]
\rangle;\qquad {\cal{O}}_{X8}=ig\langle W_{\mu\nu}[L^{\mu},L^{\nu}]
\rangle;\qquad {\cal{O}}_{X9}=ig\langle W_{\mu\nu}\tau_L\rangle \langle \tau_L[L^{\mu},L^{\nu}]
\rangle
\end{align}
are actually redundant and can be expressed in terms of oblique parameters and gauge-fermion operators. These redundancies have been known for a while~\cite{De Rujula:1991se}, but never used in phenomenological studies. In the large-$s$ limit, the oblique operators are 'projected out' and the role of the rightmost ${\cal{O}}_{X7-9}$-induced diagram in Fig.~\ref{fig:2} is equivalent to the central gauge-fermion-induced contact terms.

For neutral gauge boson pair production, the first thing to notice is that the operators we listed at NNLO are not in the form of triple-gauge corrections. However, using the equations of motion for the gauge fields one can rewrite them as 
\begin{align}
{\cal{L}}_{nTGV}&=\frac{\lambda_{ZZ}}{\Lambda^2}\partial_{\lambda}Z^{\lambda\mu}Z^{\nu}Z_{\mu\nu}+\frac{\lambda_{\gamma\gamma}}{\Lambda^2}\partial_{\lambda}F^{\lambda\mu}Z^{\nu}F_{\mu\nu}+\frac{\lambda_{Z\gamma}}{\Lambda^2}\partial_{\lambda}Z^{\lambda\mu}Z^{\nu}F_{\mu\nu}+\frac{\lambda_{\gamma Z}}{\Lambda^2}\partial_{\lambda}F^{\lambda\mu}Z^{\nu}Z_{\mu\nu}\nonumber\\
&+\frac{{\tilde{\lambda}}_{ZZ}}{\Lambda^2}\partial_{\lambda}Z^{\lambda\mu}Z^{\nu}{\tilde{Z}}_{\mu\nu}+\frac{{\tilde{\lambda}}_{\gamma\gamma}}{\Lambda^2}\partial_{\lambda}F^{\lambda\mu}Z^{\nu}{\tilde{F}}_{\mu\nu}+\frac{{\tilde{\lambda}}_{Z\gamma}}{\Lambda^2}\partial_{\lambda}Z^{\lambda\mu}Z^{\nu}{\tilde{F}}_{\mu\nu}+\frac{{\tilde{\lambda}}_{\gamma Z}}{\Lambda^2}\partial_{\lambda}F^{\lambda\mu}Z^{\nu}{\tilde{Z}}_{\mu\nu}
\end{align}
This means that the $s$-channel diagram in Fig.~\ref{fig:1} is redundant: fermionic contact terms already capture the relevant new physics in neutral triple-gauge vertices. However, as emphasized before, these operators are NNLO and their detection thus becomes rather challenging. Fortunately, final-state polarizations can be used to disentangle these suppressed new physics effects from the dominant standard model prediction. One actually finds 
\begin{align}
\frac{d\sigma_{Z_{T}^{\pm}Z_{T}^{\mp}}^j}{d\cos{\theta}}&=\frac{2\pi\alpha^2}{s}\frac{1\pm\cos\theta}{1\mp\cos\theta}\zeta_j^4;&&\qquad \frac{d\sigma_{Z_{L}Z_{T}^{\pm}}^j}{d\cos{\theta}}=-\frac{\pi\alpha^2(1\mp\cos{\theta})\zeta_j^2}{\Lambda^2} \frac{c_W{\tilde{c}}_{jW}-s_W{\tilde{c}}_{jB}}{es_{2W}}\nonumber\\
\frac{d\sigma_{\gamma^{\mp} Z_{T}^{\pm}}^j}{d\cos\theta}&=\frac{2\pi\alpha^2}{s}\frac{1\pm \cos{\theta}}{1\mp \cos{\theta}}\zeta_j^2;&&\qquad
\frac{d\sigma_{\gamma^{\pm} Z_L}^j}{d\cos\theta}=-\frac{\pi\alpha^2(1\mp \cos{\theta})\zeta_j}{\Lambda^2} \frac{s_W{\tilde{c}}_{jW}+c_W{\tilde{c}}_{jB}}{es_{2W}}
\end{align}
which shows that, although suppressed, the LT channel turns out to be a remarkably clean probe of anomalous neutral triple-gauge vertices. 

\section{Conclusions}
The systematics of the EFT description of a strongly-coupled EWSB have been reviewed and applied to both charged and neutral boson pair production. I have shown that a full-fledged EFT analysis in the regime $v^2\ll s\ll \Lambda^2$ allows general new physics effects (including anomalous triple-gauge vertices) to be entirely encoded in gauge-fermion corrections. This applies to both charged and neutral pair production as a consequence of straightforward application of the equations of motion. However, their phenomenological signatures are rather distinct: in $WW$ production, new physics effects are $s$-enhanced with respect to the standard model cross section through the LL final-state polarizations. In contrast, for $ZZ$ and $\gamma Z$ production, new physics operators are extremely suppressed in the unpolarized cross section but neatly dominate the LT final-state polarizations. In both cases, the effects can easily reach the $20\%$ corrections in the unpolarized and LT-polarized cross sections, respectively, for energies in the range $\sqrt{s}\sim (0.6-1)$ TeV.  
 

\end{document}